\documentclass[twocolumn]{aastex63}
\usepackage{float}
\usepackage[normalem]{ulem}

\begin{document}

\title{Parallax Systematics and Photocenter Motions of Benchmark Eclipsing Binaries in {\it Gaia\/} EDR3}
\author[0000-0002-3481-9052]{Keivan G.\ Stassun}
\affiliation{Vanderbilt University, Department of Physics \& Astronomy, 6301 Stevenson Center Ln., Nashville, TN  37235, USA}
\author{Guillermo Torres}
\affiliation{Harvard-Smithsonian Center for Astrophysics, 60 Garden St., Cambridge, MA 02138, USA}

\begin{abstract}
Previous analyses of various standard candles observed by the {\it Gaia\/} satellite have reported statistically significant systematics in the parallaxes that have improved from $\sim$250~$\mu$as in the first data release (DR1) to 50--80~$\mu$as in the second data release (DR2). 
Here we examine the parallaxes newly reported in the {\it Gaia\/} early third data release (EDR3) using the same sample of benchmark eclipsing binaries (EBs) we used to assess the DR1 and DR2 parallaxes. 
We find a mean offset of $-37\pm20$~$\mu$as ({\it Gaia\/}$-$EB), which decreases to $-15\pm18$~$\mu$as after applying the corrections recommended by the {\it Gaia\/} Mission team; global systematics in the {\it Gaia\/} parallaxes have clearly improved and are no longer statistically significant for the EB sample, which spans $5\lesssim{G}\lesssim12$ in brightness and 0.03--3~kpc in distance.
We also find that the RUWE goodness-of-fit statistic reported in {\it Gaia}~DR3 is highly sensitive to unresolved companions (tertiaries in the case of our EB sample) as well as to photocenter motion of the binaries themselves. RUWE is nearly perfectly correlated ($r^2=0.82$) with photocenter motions down to $\lesssim$0.1~mas, and surprisingly this correlation exists entirely within the nominal ``good" RUWE range of 1.0--1.4. This suggests that RUWE values even slightly greater than 1.0 may signify unresolved binaries in {\it Gaia}, and that the RUWE value can serve as a quantitative predictor of the photocenter motion.\\
\end{abstract}

\section{Introduction}\label{sec:intro}
The trigonometric parallaxes for $\sim$10$^9$ stars from the {\it Gaia\/} mission have heralded a new era of precision astrophysics for stars, exoplanets, and the Milky Way. 
It is essential to assess these parallaxes for potential biases, particularly in light of the experience from {\it Hipparcos\/}, which suffered a significant bias in at least the case of the Pleiades cluster \citep[e.g.,][]{Pinsonneault:1998}. 
Such a check requires a set of benchmark stars whose parallaxes are determined independent of the {\it Gaia\/} parallaxes, and preferably independent of parallaxes altogether.

To this end, 
\citet{Stassun:2016} assembled a sample of 158 eclipsing binary stars (EBs) whose radii and effective temperatures are known empirically and precisely, such that their bolometric luminosities are determined to high precision (via the Stefan-Boltzmann relation) and therefore independent of assumed distance. 
\citet{Stassun:2016} also measured the bolometric fluxes for these EBs which, together with the precisely known bolometric luminosities, yielded the EB distances. 
%
While the precision on the predicted EB parallaxes is $\approx$190~$\mu$as on average, 
the EB sample is large enough that it should be possible in principle to assess average systematics down to $\sim 190/\sqrt{158}\sim 15$~$\mu$as.

In \citet{StassunTorres:2016} we reported an initial assessment of the {\it Gaia\/} first data release (DR1) parallaxes, finding a significant average offset of $-250\pm 50$~$\mu$as, in the sense that the {\it Gaia\/} DR1 parallaxes were too small. 
That finding, which was consistent with the expected systematic error floor of 300~$\mu$as for {\it Gaia\/} DR1 \citep{Gaia:2016}, was corroborated by other authors on the basis of ground-based parallaxes of nearby M~dwarfs \citep{Jao:2016}, asteroseismic stellar radii in the Kepler field \citep{Huber:2017},
and stellar radii from granulation ``flicker" \citep{Stassun:2018}. 
We reprised that analysis for the {\it Gaia\/} DR2 parallaxes in \citet{StassunTorres:2018}, finding a much smaller offset of about $-80$~$\mu$as, comparable to the offsets averaging about $-60$~$\mu$as found by a number of other studies \citep[e.g.,][]{Riess:2018,Zinn:2018,Kounkel:2018,Arenou:2018}. 

In this paper we apply the same benchmark EB sample of \citet{Stassun:2016} to the {\it Gaia\/} early third data release (EDR3). 
Section~\ref{sec:data} summarizes the EB and {\it Gaia\/} data used. Section~\ref{sec:results} presents the key finding that we no longer observe a statistically significant global offset in the {\it Gaia\/} parallaxes. 
Section~\ref{sec:disc} discusses the implications of a correlation that we find between photocenter motion and {\it Gaia\/} reported goodness-of-fit statistics for the identification of unresolved binaries throughout the {\it Gaia\/} catalog. Section~\ref{sec:summary} concludes with a summary of our findings.

\section{Data}\label{sec:data}
We adopted the predicted parallaxes for the 151 EBs from \citet{Stassun:2016} that were also included in our previous analysis of the {\it Gaia\/} DR2 parallaxes \citep{StassunTorres:2018}, 
all of which have parallaxes available in {\it Gaia\/} EDR3.
The EBs are all bright, with {\it Gaia\/} magnitudes in the range $5\lesssim{G}\lesssim12$, and extend out to $\sim$3~kpc, with parallaxes in the range $\pi\approx\,$0.3--30~mas.
In addition, all of the EBs have distance uncertainties better than 15\%, and thus the choice of prior on the EB distance should not be important for inferring their parallaxes \citep[e.g.,][]{Bailer-Jones:2016}.

\citet{Stassun:2016} identified 32 of the EBs as having known or suspected tertiary companions from the original EB analysis and/or from subsequent followup studies. We flag those systems separately in our analysis as part of our examination of the effects of such companions. 
To that end, we use the Renormalised Unit Weight Error (RUWE) statistic reported in {\it Gaia\/} EDR3 as the primary astrometric goodness-of-fit indicator.

\section{Results}\label{sec:results}

\begin{figure}[!ht]
\centering
\includegraphics[width=\linewidth,trim=10 0 20 70,clip]{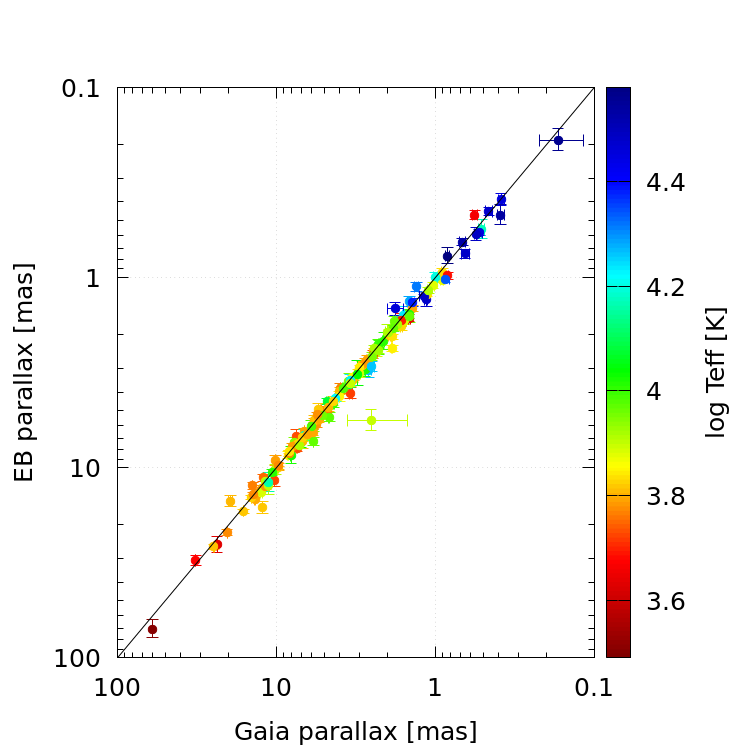}
\includegraphics[width=\linewidth,trim=13 0 35 60,clip]{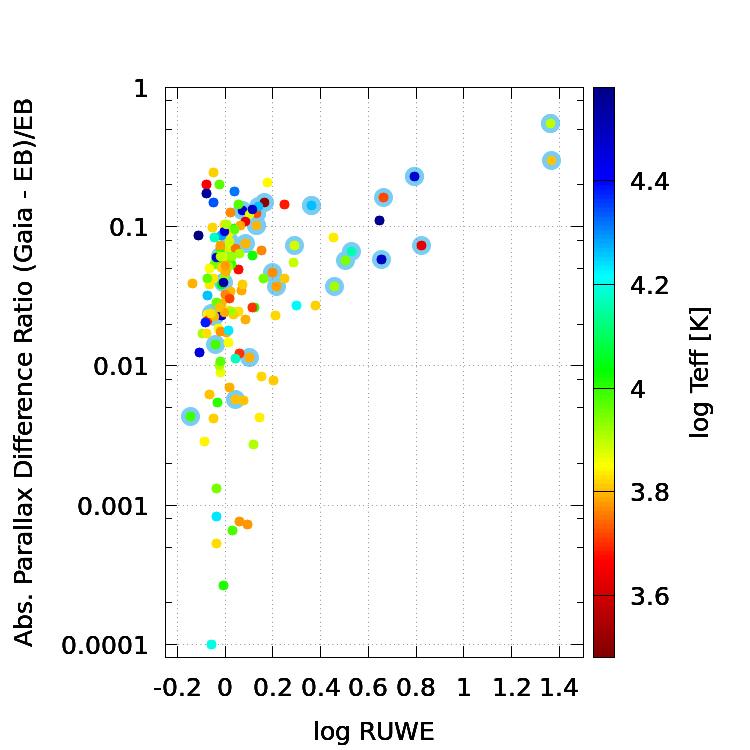}
\caption{{\it (Top:)} Direct comparison of predicted parallaxes from the eclipsing binary sample of \citet[][$\pi_{\rm EB}$]{Stassun:2016} versus the parallaxes from the {\it Gaia\/} early third data release ($\pi_{\rm Gaia}$). The one-to-one line is shown in black, and color represents $T_{\rm eff}$ for the EBs as a proxy for system color. 
{\it (Bottom:)} The absolute fractional parallax difference $\left|\left(\pi_{\rm Gaia}-\pi_{\rm EB}\right)/\pi_{\rm EB}\right|$ versus the {\it Gaia\/} astrometric goodness-of-fit statistic (RUWE). Blue haloes identify EBs known or suspected to have unresolved tertiary companions that could affect the {\it Gaia\/} astrometric solution; indeed, the upper-right portion of the diagram, representing EBs with large parallax differences relative to {\it Gaia\/} and exhibiting large RUWE in {\it Gaia\/}, is populated almost entirely by these systems.}
\label{fig:pxvspx} 
\end{figure}

Figure~\ref{fig:pxvspx} (top) shows the direct comparison of the EB parallax predictions from \citet{Stassun:2016} versus the {\it Gaia\/} EDR3 parallaxes for the study sample. 
To first order, the agreement between the {\it Gaia\/} EDR3 and EB parallaxes appears excellent. 

A closer examination of the differences between the {\it Gaia\/} parallaxes ($\pi_{\rm Gaia}$) and the EB parallaxes ($\pi_{\rm EB}$) shows that while the vast majority agree to within a few percent, the differences range from as small as 0.01\% in the best cases to as large as $\sim$50\% in the worst cases (Figure~\ref{fig:pxvspx}, bottom). Importantly, we observe that the parallax differences are very strongly correlated with the {\it Gaia\/} RUWE goodness-of-fit statistic, and that the sources with RUWE$>$1.4 tend to be those with known or suspected tertiaries (points with blue haloes in the upper-right of Figure~\ref{fig:pxvspx}, bottom). A Kendall's $\tau$ test yields a probability of 0.0004 for the null hypothesis of no correlation between parallax difference and RUWE. And while the correlation does not appear to be simply linear, a simple linear correlation alone can account for 56\% of the variance between the two quantities (i.e., $r^2=0.56$). 

\begin{figure}[!ht]
    \centering
    \includegraphics[width=\linewidth,trim=10 0 40 80,clip]{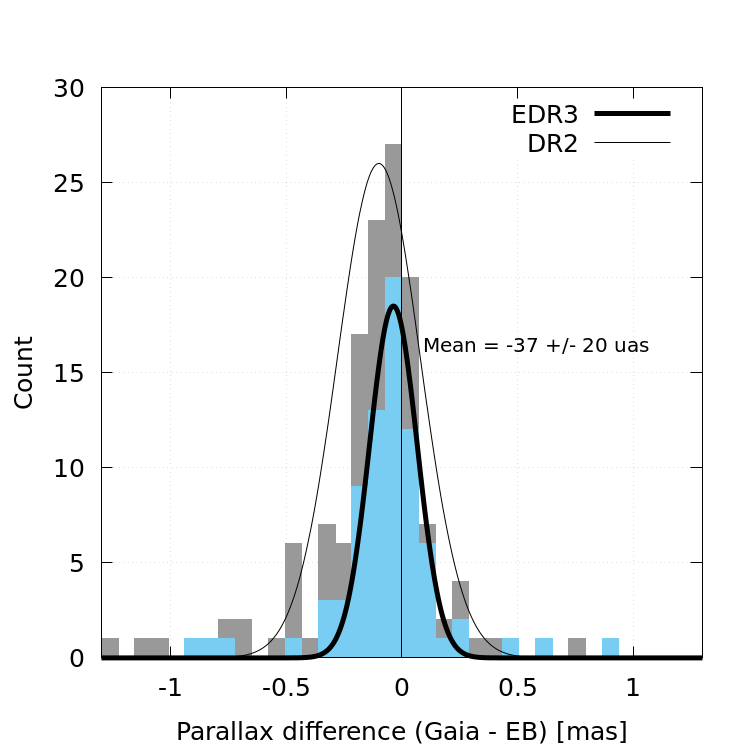}
    \caption{Distribution of $\Delta\pi$ ({\it Gaia\/}$-$EB) for the entire EB sample (gray histogram) 
    and for the subset we use to evaluate the global offset (blue histogram; see the text), with a Gaussian fit 
    to the latter (thick black curve).
    The thin black curve represents the Gaussian fit to the $\Delta\pi$ distribution that we found for DR2 \citep{StassunTorres:2018}, scaled to the height of the gray histogram for visual comparison. 
    }
    \label{fig:delta_plx_hist}
\end{figure}

Figure~\ref{fig:delta_plx_hist} presents the overall distribution of parallax differences in the sense of $\pi_{\rm Gaia}-\pi_{\rm EB}$. 
The distribution appears roughly symmetric and normally distributed, with perhaps a sharper peak and more extended wings than a Gaussian. 
In comparison to the distribution of parallax differences that we found in our DR2 analysis \citep[][thin black curve in Figure~\ref{fig:delta_plx_hist}]{StassunTorres:2018}, the distribution now is markedly narrower, indicating that the scatter in the {\it Gaia\/} EDR3 parallaxes is improved relative to the DR2 parallaxes.

Based on the finding above that the RUWE statistic is correlated with the presence of a tertiary companion and thus with larger parallax differences, we only consider the subset of EBs with RUWE~$<$~1.4 and that are not known or suspected of possessing tertiaries. 
We moreover consider only those with photocenter motions smaller than 0.1~mas (see Section~\ref{sec:photo}), 
leaving 76 EBs as our best sample to assess the global parallax offset (blue histogram in Figure~\ref{fig:delta_plx_hist}). The mean offset from a simple (unweighted) Gaussian fit is $\Delta\pi = -37 \pm 20$~$\mu$as, where the quoted uncertainty is the standard error of the mean (i.e., $\sigma_{\rm r.m.s.}/\sqrt{76}$). 

\citet{Lindegren:2020} have provided a prescription for small parallax corrections as a function of brightness, color, and position on the sky; those corrections average $\sim$20~$\mu$as for the EB sample, and applying them (on a star-by-star basis) yields a corrected mean parallax offset of $\Delta\pi=-15\pm18$~$\mu$as
(not shown in Figure~\ref{fig:delta_plx_hist}), calculated as above. 
These corrections thus 
appear promising.

\section{Discussion}\label{sec:disc}

\subsection{Accuracy of the Eclipsing Binary Parallaxes}

As of the {\it Gaia\/} early third data release (EDR3), we no longer observe a statistically significant global offset in the {\it Gaia\/} parallaxes relative to the sample of benchmark eclipsing binaries from \citet{Stassun:2016}. The analysis of Section~\ref{sec:results} suggests an upper limit to any remaining global offset of $-33$~$\mu$as (in the sense of {\it Gaia\/}~$-$~EB), after adjusting the {\it Gaia\/} EDR3 parallaxes by the small corrections prescribed by \citet{Lindegren:2020}. Even without those corrections, our finding here of no significant global offset continues the trend of improvement in the {\it Gaia\/} parallaxes, from $-250$~$\mu$as in DR1 to $-80$~$\mu$as in DR2 to $-37$~$\mu$as in EDR3. This latest estimate is now comparable to the fundamental accuracy attainable by the EB sample (see Section~\ref{sec:intro}); thus, absent a major enlargement of the benchmark EB sample, it appears that comparisons of the {\it Gaia\/} parallaxes to these EBs have reached the limit of their utility. 

At the same time, the steady improvement of the {\it Gaia\/} parallaxes relative to the EB sample down to the nominal limits of the EB accuracy serves as a validation of certain assumptions built in to the EB sample, in particular the effective temperature scale. 
As we noted in our previous analyses of the {\it Gaia\/} DR1 and DR2 parallaxes \citep{StassunTorres:2016,StassunTorres:2018}, in principle the offsets we reported could have been due to systematics in 
the EB $T_{\rm eff}$ values: unlike the EB stellar radii, for example, which are determined from simple geometry, the $T_{\rm eff}$ values are determined from spectral analysis and/or spectral typing and/or color relations. 
For example, the global parallax offset of $-80$~$\mu$as that we reported in \citet{StassunTorres:2018} would 
require a systematic error in $T_{\rm eff}$ of 
$\sim$50~K 
in the sense 
that the EBs would have to be systematically too cool.
%
This possibility was 
discounted by \citet{StassunTorres:2018} for multiple reasons, 
including comparison to the {\it Hipparcos\/} parallaxes \citep{Stassun:2016}
and the fact that the various EBs in our study sample have had their $T_{\rm eff}$ determined by different methodologies and different calibrations, 
making it very unlikely that any individual biases should produce a net systematic offset in a large sample of EBs spanning a large range of $T_{\rm eff}$. 
 
\subsection{Unresolved Companions and Photocenter Motion}\label{sec:photo}

We observed in Figure~\ref{fig:pxvspx} (bottom) that the {\it Gaia\/} RUWE astrometric goodness-of-fit statistic is clearly sensitive to the presence of unresolved companions (tertiaries, in the case of the EB sample). Nearly all of the EBs with RUWE~$>$~1.4 \citep[typically used as the threshold suggesting a poor astrometric solution; see, e.g.,][]{Lindegren:RUWE} are previously known or suspected triples \citep[see][and references therein]{Stassun:2016}. Consequently, we suggest that the few EBs with RUWE~$>$~1.4 that are not flagged as triples in Figure~\ref{fig:pxvspx} (bottom) are in fact triples but have simply not been previously identified as such. 
Our finding that the RUWE is sensitive to the presence of unresolved companions is consistent with the results of \citet{Belokurov:2020} for {\it Gaia\/} DR2.

\begin{figure}[!ht]
    \centering
    \includegraphics[width=\linewidth,trim=10 5 35 85,clip]{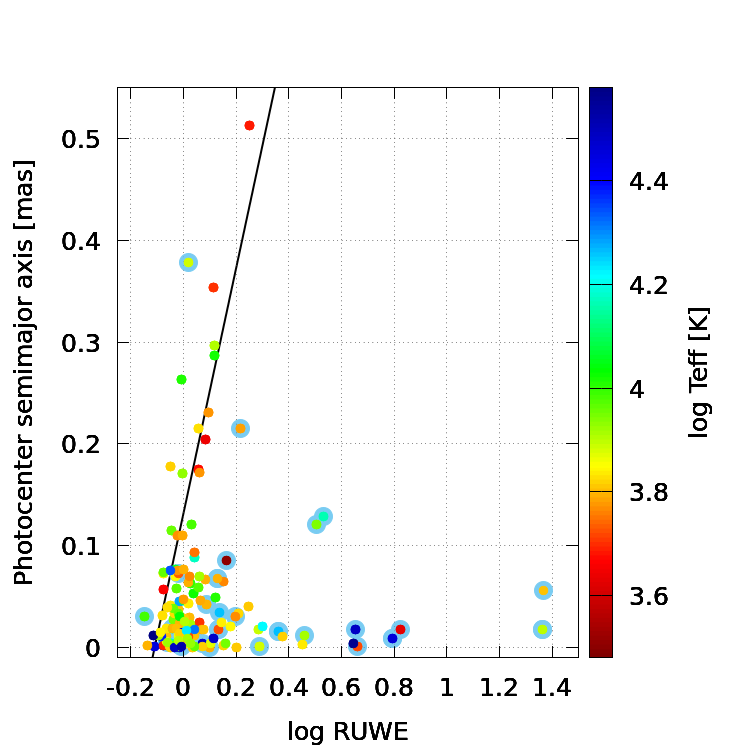}
    \caption{Same as Figure~\ref{fig:pxvspx} (bottom), except the ordinate now represents the predicted photocenter motion of the EBs. The solid black line represents a linear least-squares fit of photocenter motion versus $\log_{10}\,$RUWE for EBs with RUWE$\,<\,$1.4, photocenter semimajor axis $\ge\,$0.1~mas, and that are not known or suspected of possessing tertiary companions (blue haloes).}
    \label{fig:ruwe_vs_photo}
\end{figure}

Interestingly, and somewhat surprisingly, we find that RUWE values below 1.4 are not devoid of astrophysical information. As shown in Figure~\ref{fig:ruwe_vs_photo}, RUWE values in the range 1.0--1.4 are very strongly correlated with the predicted photocenter motions of the EBs. 
The photocenter semimajor axis of a binary in angular units is
related to the angular semimajor axis of the orbit by $a_{\rm
phot}\arcsec = a\arcsec (B-\beta)$ \citep[e.g.,][]{vandeKamp:1967}, in
which $B =M_2/(M_1+M_2)$ and
$\beta = \ell_2/(\ell_1+\ell_2)$ are the secondary's fractional mass and fractional light at the
wavelength of the observation (the {\it Gaia\/} $G$ band). With the
approximation that the fractional light in $G$ is equal to the
fractional bolometric luminosity $L_2/(L_1+L_2)$, we computed the latter with the
Stefan-Boltzmann law using the known effective temperatures and radii
of each EB component. $B$ follows from the known masses, and $a\arcsec$
from Kepler's third law with the masses and {\it Gaia\/} parallax, and the known
orbital periods.

For EBs with RUWE$\,<\,$1.4, photocenter semimajor axis $a_{\rm phot}\arcsec \ge 0.1$~mas, and that are not known or suspected of possessing tertiary companions, we find that a simple linear correlation accounts for fully 82\% of the variance ($r^2=0.82$), and a Kendall's $\tau$ test returns a probability of 0.0001 for the null hypothesis of no correlation. 
The correlation is of the form
$$a_{\rm phot}\arcsec\,{\rm [mas]}=1.204\times\log_{10}{\rm RUWE}+0.13$$
for $1.0\le{\rm RUWE}\le1.4$.

\section{Summary and Conclusions}\label{sec:summary}

We have found that the systematic offsets previously reported in the average zero-point of the {\it Gaia\/} DR1 and DR2 parallax measurements are no longer statistically significant in the {\it Gaia\/} early third data release (EDR3). 
Formally, the parallax offset that we measure is $-15\pm18$~$\mu$as (or an upper limit of $-33$~$\mu$as) after applying the small parallax corrections prescribed by \citet{Lindegren:2020}. 
The reference for this determination is the set of independently inferred parallaxes from the \citet{Stassun:2016} benchmark sample of well-studied eclipsing binaries with a wide range of brightnesses ($5<G<12$), distances up to 3~kpc, and distributed over the entire sky. 

That the systematic offset we measured has steadily improved from $-250$~$\mu$as in {\it Gaia\/} DR1 \citep{StassunTorres:2016} is a testament to the quality of the vetting of the {\it Gaia\/} Mission and an affirmation of the fundamental accuracy of the benchmark eclipsing binaries, whose predicted parallaxes have remained unchanged. At the same time, any remaining systematics in the {\it Gaia\/} parallaxes are now at or below the fundamental accuracy limit of the benchmark eclipsing binary sample. 
Other standard candles such as RR Lyrae variables can provide additional constraints \citep[e.g.,][]{Bhardwaj:2021}. 

Finally, we find that the {\it Gaia\/} RUWE astrometric goodness-of-fit statistic is highly sensitive to unresolved companions (tertiaries in the case of the eclipsing binaries) as well as to photocenter motion of the binaries themselves. In fact, setting aside the EBs with tertiary companions, RUWE is very strongly correlated ($r^2=0.82$) with photocenter motions down to $\lesssim\,$0.1~mas, and remarkably this correlation exists entirely within the nominal ``good" RUWE range of 1.0--1.4. This somewhat surprising finding implies that RUWE values even slightly greater than 1.0 may in general signify unresolved binaries throughout the {\it Gaia\/} catalog, and the RUWE value can serve as a quantitative predictor of the photocenter motion.

\section*{Acknowledgments}
This work has made use of the Filtergraph data visualization service \citep{Burger:2013}. 
We thank the anonymous referee for the helpful comments.
K.G.S.\ acknowledges partial support from NASA grant 17-XRP17 2-0024. G.T.\ acknowledges partial support from NSF grant AST-1509375.
This work has made use of data from the European Space Agency (ESA) mission {\it Gaia\/} (http://www.cosmos.esa.int/gaia), processed by the {\it Gaia\/} Data Processing and Analysis Consortium (DPAC, http://www.cosmos.esa.int/web/gaia/dpac/consortium). Funding for the DPAC has been provided by national institutions, in particular the institutions participating in the {\it Gaia\/} Multilateral Agreement.


\begin{thebibliography}{}

\bibitem[Arenou et al.(2018)]{Arenou:2018} Arenou, F., Luri, X., Babusiaux, C., et al.\ 2018, \aap, 616, A17. doi:10.1051/0004-6361/201833234

\bibitem[Bailer-Jones(2015)]{Bailer-Jones:2016} Bailer-Jones, C.~A.~L.\ 2015, \pasp, 127, 994 



\bibitem[Belokurov et al.(2020)]{Belokurov:2020} Belokurov, V., Penoyre, Z., Oh, S., et al.\ 2020, \mnras, 496, 1922. doi:10.1093/mnras/staa1522

\bibitem[Bhardwaj et al.(2021)]{Bhardwaj:2021}
Bhardwaj, A., et al.\ 2021, arxiv:2012.13495

\bibitem[Burger et al.(2013)]{Burger:2013} Burger, D., Stassun, K.~G., Pepper, J., et al.\ 2013, Astronomy and Computing, 2, 40







\bibitem[Gaia Collaboration et al.(2016)]{Gaia:2016} Gaia Collaboration, Brown, A.~G.~A., Vallenari, A., et al.\ 2016, \aap, 595, A2. doi:10.1051/0004-6361/201629512



\bibitem[Huber et al.(2017)]{Huber:2017} Huber, D., Zinn, J., Bojsen-Hansen, M., et al.\ 2017, \apj, 844, 102 

\bibitem[Jao et al.(2016)]{Jao:2016} Jao, W.-C., Henry, T.~J., Riedel, A.~R., et al.\ 2016, \apjl, 832, L18 



\bibitem[Kounkel et al.(2018)]{Kounkel:2018} Kounkel, M., Covey, K., Su{\'a}rez, G., et al.\ 2018, \aj, 156, 84. doi:10.3847/1538-3881/aad1f1




\bibitem[Lindegren et al.(2020)]{Lindegren:2020} Lindegren, L., Bastian, U., Biermann, M., et al.\ 2020, arXiv:2012.01742

\bibitem[Lindegren(2018)]{Lindegren:RUWE}
Lindegren, L.\ 2018, Gaia Technical Note GAIA-C3-TN-LU-LL-124-01. Obtained from https://www.cosmos.esa.int/web/gaia/public-dpac-documents 


\bibitem[Pinsonneault et al.(1998)]{Pinsonneault:1998} Pinsonneault, M.~H., Stauffer, J., Soderblom, D.~R., King, J.~R., \& Hanson, R.~B.\ 1998, \apj, 504, 170 




\bibitem[Ricker et al.(2015)]{Ricker:2011} Ricker, G.~R., Winn, J.~N., Vanderspek, R., et al.\ 2015, Journal of Astronomical Telescopes, Instruments, and Systems, 1, 014003 

\bibitem[Riess et al.(2018)]{Riess:2018} Riess, A.~G., Casertano, S., Yuan, W., et al.\ 2018, \apj, 861, 126. doi:10.3847/1538-4357/aac82e

\bibitem[Stassun et al.(2018)]{Stassun:2018} Stassun, K.~G., Corsaro, E., Pepper, J.~A., \& Gaudi, B.~S.\ 2018, \aj, 155, 22 


\bibitem[Stassun \& Torres(2016a)]{Stassun:2016} Stassun, K.~G., \& Torres, G.\ 2016a, \aj, 152, 180 

\bibitem[Stassun \& Torres(2016b)]{StassunTorres:2016} Stassun, K.~G., \& Torres, G.\ 2016b, \apjl, 831, L6 

\bibitem[Stassun \& Torres(2018)]{StassunTorres:2018} Stassun, K.~G. \& Torres, G.\ 2018, \apj, 862, 61. doi:10.3847/1538-4357/aacafc



\bibitem[van de Kamp(1967)]{vandeKamp:1967} van de Kamp, P. 1967,
Principles of Astrometry (San Francisco, CA: Freeman), p.\ 168


\bibitem[Zinn et al.(2019)]{Zinn:2018} Zinn, J.~C., Pinsonneault, M.~H., Huber, D., et al.\ 2019, \apj, 878, 136. doi:10.3847/1538-4357/ab1f66

\end{thebibliography}
\end{document}